\documentclass[aip,,twocolumn,jmp,amsmath,amssymb,reprint]{revtex4-1}
\usepackage{graphicx}
\usepackage{dcolumn}
\usepackage{float}
\usepackage{caption}
\usepackage{subcaption}
\usepackage{color}
\usepackage[normalem]{ulem}
\usepackage{booktabs}
\begin{document}


\title{EUV induced defects on few-layer graphene} 

\author{A. Gao}
\email{a.gao@differ.nl.}
 \affiliation{FOM-Dutch Institute for Fundamental Energy Research, Edisonbaan 14,3439 MN Nieuwegein, the Netherlands.}
\author{P.J. Rizo}%
\affiliation{ASML, De Run 6501, 5504DR Veldhoven, the Netherlands.}
\author{E. Zoethout}
 \affiliation{FOM-Dutch Institute for Fundamental Energy Research, Edisonbaan 14,3439 MN Nieuwegein, the Netherlands.} 
\author{L. Scaccabarozzi}
\affiliation{ASML, De Run 6501, 5504DR Veldhoven, the Netherlands.}
\author{C.J. Lee}
 \affiliation{FOM-Dutch Institute for Fundamental Energy Research, Edisonbaan 14,3439 MN Nieuwegein, the Netherlands.}
\author{V. Banine}
 \affiliation{ASML, De Run 6501, 5504DR Veldhoven, the Netherlands.}
\author{F. Bijkerk}
 \affiliation{FOM-Dutch Institute for Fundamental Energy Research, Edisonbaan 14,3439 MN Nieuwegein, the Netherlands.}
 \affiliation{MESA+ Institute for Nanotechnology, PO Box 217, University of Twente, 7500 AE, Enschede, the Netherlands.}
\date{\today}

\begin{abstract}
We use Raman spectroscopy to show that exposing few-layer graphene to extreme ultraviolet (EUV, 13.5~nm) radiation, i.e. relatively low photon energy, results in an increasing density of defects. Furthermore, exposure to EUV radiation in a H$_2$ background increases the graphene dosage sensitivity, due to reactions caused by the EUV induced hydrogen plasma. X-ray photoelectron spectroscopy (XPS) results show that the sp$^2$ bonded carbon fraction decreases while the sp$^3$ bonded carbon and oxide fraction increases with exposure dose. Our experimental results confirm that even in reducing environment oxidation is still one of the main source of inducing defects. 
\end{abstract}


\maketitle 

\section{\label{sec:intro}Introduction}

Graphene is a single planar sheet of sp$^2$ bonded carbon atoms which are closely packed in a honeycomb-like crystal structure. It is the basis of many carbon-based materials, e.g., stacked into graphite, rolled into carbon nanotubes or wrapped into buckyballs~\cite{geim2007rise,geim2009graphene,novoselov2005two}.
Graphene has unique physical properties, such as quantum electronic transport, a tunable band gap, extremely high mobility, high elasticity, and electromechanical modulation~\cite{novoselov2005two,zhang2005experimental,han2007energy,bolotin2008ultrahigh,lee2008measurement,bunch2007electromechanical}. This makes graphene a promising material for many applications, including graphene transistors, electronic circuits, and solar cells, as well as other applications in biology and chemistry~\cite{novoselov2005two,zhang2005experimental,han2007energy,bolotin2008ultrahigh,lee2008measurement,bunch2007electromechanical}. However, one of the key requirements for such applications is the control of defects, such as vacancies, dislocations or adatoms. The electronic properties of graphene are greatly affected by the presence of defects because they can act as scattering centers for electrons, reducing sheet conductivity~\cite{boukhvalov2008chemical}. Defects associated with dangling bonds can enhance the chemical reactivity of graphene~\cite{cortijo2007effects,rutter2007scattering}. Likewise, the presence of defects reduces the thermal conductivity of graphene~\cite{hao2011mechanical}.

The unique properties make graphene an attractive candidate for applications in radiation-rich environment. However the presence of defect may affect its performance. Therefore, it is critical to understand the radiation-induced damage in graphene. Zhou \textit{et al}~\cite{zhou2009instability}, reported that soft x-rays can easily break the sp$^2$ bond structure and form defects in graphene that is weakly bound to the substrate. Hicks \textit{et al}~\cite{hicks2011x} also studied multilayer graphene, grown on SiC, before and after 10~keV x-ray irradiation in air. They concluded that defects were generated due to surface etching by reactive oxygen species created by x-rays. In this paper, we focus on  defect generation in graphene, induced by exposure to 13.5~nm (EUV) radiation under a variety of background conditions. We compare the rate at which defects are induced by EUV in a vacuum condition, and the rate at which defects are induced by exposure to EUV in a background of molecular hydrogen. We show that, defects are introduced in both cases, though at different rates. Surprisingly, our data also show that, even in a reducing environment, oxidation is still one of the main sources of EUV induced defects. The experimental results are important for illustrating the damage-creating mechanisms upon photon interaction as well as designing graphene-based components for EUV lithography systems.

\begin{table*}[!htb]
\caption{\label{tab:table1}Experimental settings summary. Two parameters vary among different experiments: exposure time to EUV radiation and/or H$_2$, hydrogen pressure.} 
\begin{ruledtabular}
\begin{tabular}{lccc}
Sample&$S_{ref}$&$S_{EUV}$&$S_{EUV+H_2}$\\
 \hline
Exposure time~(hr)&NA&8&8\\
H$_2$ pressure~(mbar)&NA&0&$5\times10^{-2}$\\
Chamber pressure~(mbar)&NA&$1\times10^{-8}$&$1\times10^{-8}$\\
\end{tabular}
\end{ruledtabular}
\end{table*}

\section{\label{sec:exp}Experiments}
Graphene samples in this report were produced by the Graphene Supermarket. A few layers of graphene were grown on $25\times25$mm$^2$ Ni/Si substrate with chemical vapor deposition method. The number of layers of graphene varies from 1 to 7, with an average of 4 over the sample. Three groups of experiments were performed: 1) a pristine sample served as a reference (refer to $S_{ref}$) and was not exposed; 2) a sample was exposed to EUV irradiation ($S_{EUV}$) without molecular hydrogen in the background gas; 3) a sample was exposed to EUV irradiation in a $5\times10^{-2}$ mbar H$_2$ background ($S_{EUV+H_2}$). The other experimental settings are summarized in table~\ref{tab:table1}.
\begin{figure}[!htb]
\centering
\includegraphics[width=0.45\textwidth]{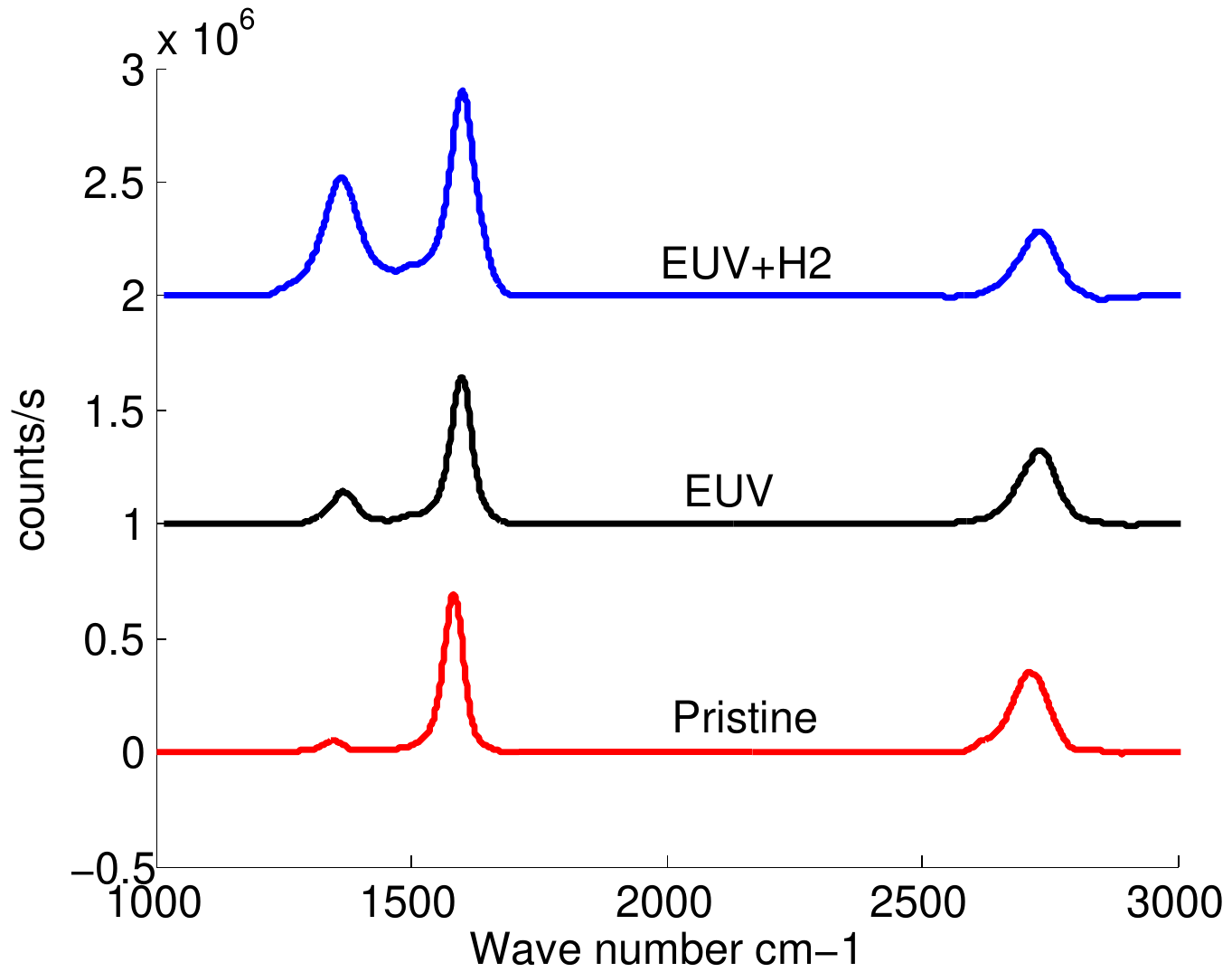}
\caption{\label{fig:spectra} (color online) Comparison of the whole Raman spectra and The spectrum for the example exposed to EUV in a hydrogen background has the highest D peak intensity. The spectra for the samples exposed to d EUV irradiation show slightly lower D peak intensity. The pristine sample has the lowest D peak intensity. Note that the spectra are separated by an offset of $5\times10^5$ counts/s.}
\end{figure}
Graphene samples were irradiated by an EUV source (Philips EUV Alpha Source 2) with a repetition rate of 1~kHz and an average dose of 0.1~mJ/cm$^2$ per pulse. Raman spectra were collected with a home-built system. In this system, a 532~nm diode-pumped solid state laser is used to excite the samples with an illumination spot of $3.5\times0.1$~mm$^2$ and a power density of 200W/cm$^2$. The collection efficiency of the detector system was calibrated using the HG-1 Mecury Argon Calibration Light Source and AvaLight-D(H)-S Deuterium-Halogen Light Source. 2D Raman intensity maps were acquired by collecting Raman signal over the central $2\times0.1$~mm$^2$ area. The transverse distance between two data points was set to 500~$\mu$m, and along the longitudinal direction, the data points were collected continuously. XPS was measured by monochromatic Al-Kalpha, Thermo Fisher Theta probe with a footprint of 1mm diameter.

\section{\label{sec:results}Results and Discussion}
\subsection{Raman analysis}

\begin{figure*}[!htb]
        \begin{subfigure}{0.25\textwidth}
                \centering
                \includegraphics[width=\textwidth]{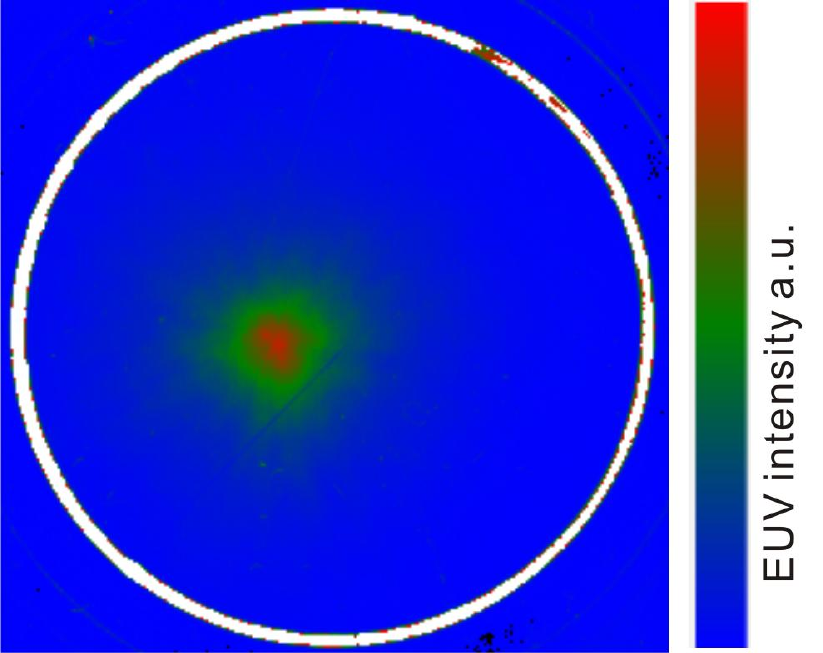}
                \caption{EUV Intensity profile}
               \label{fig:euvprof}                
        \end{subfigure}~
        \begin{subfigure}{0.25\textwidth}
                \centering
                \includegraphics[width=\textwidth]{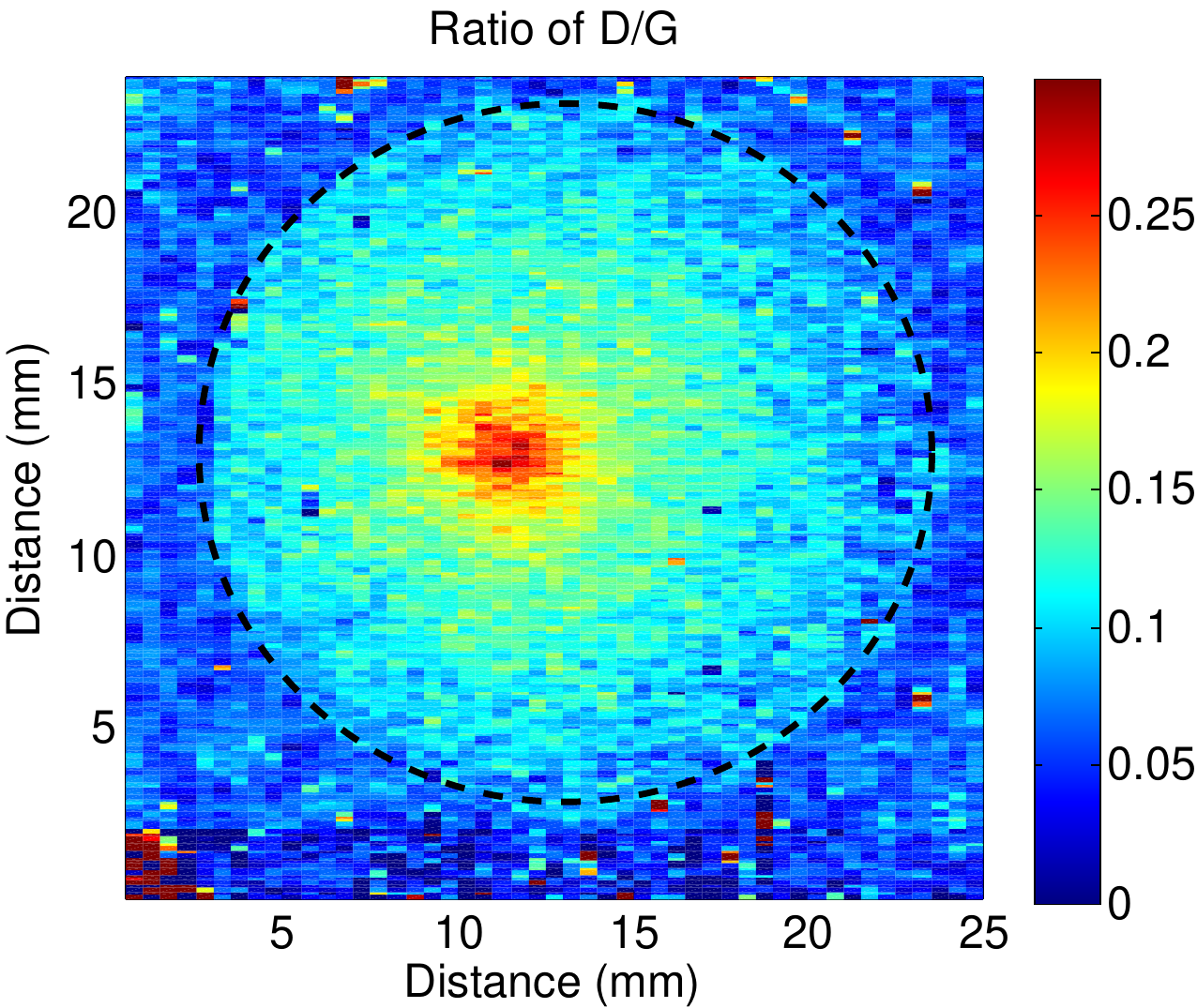}
                \caption{EUV only}
                \label{fig:EUV}
        \end{subfigure}%
        ~ 
        \begin{subfigure}{0.25\textwidth}
                \centering
                \includegraphics[width=\textwidth]{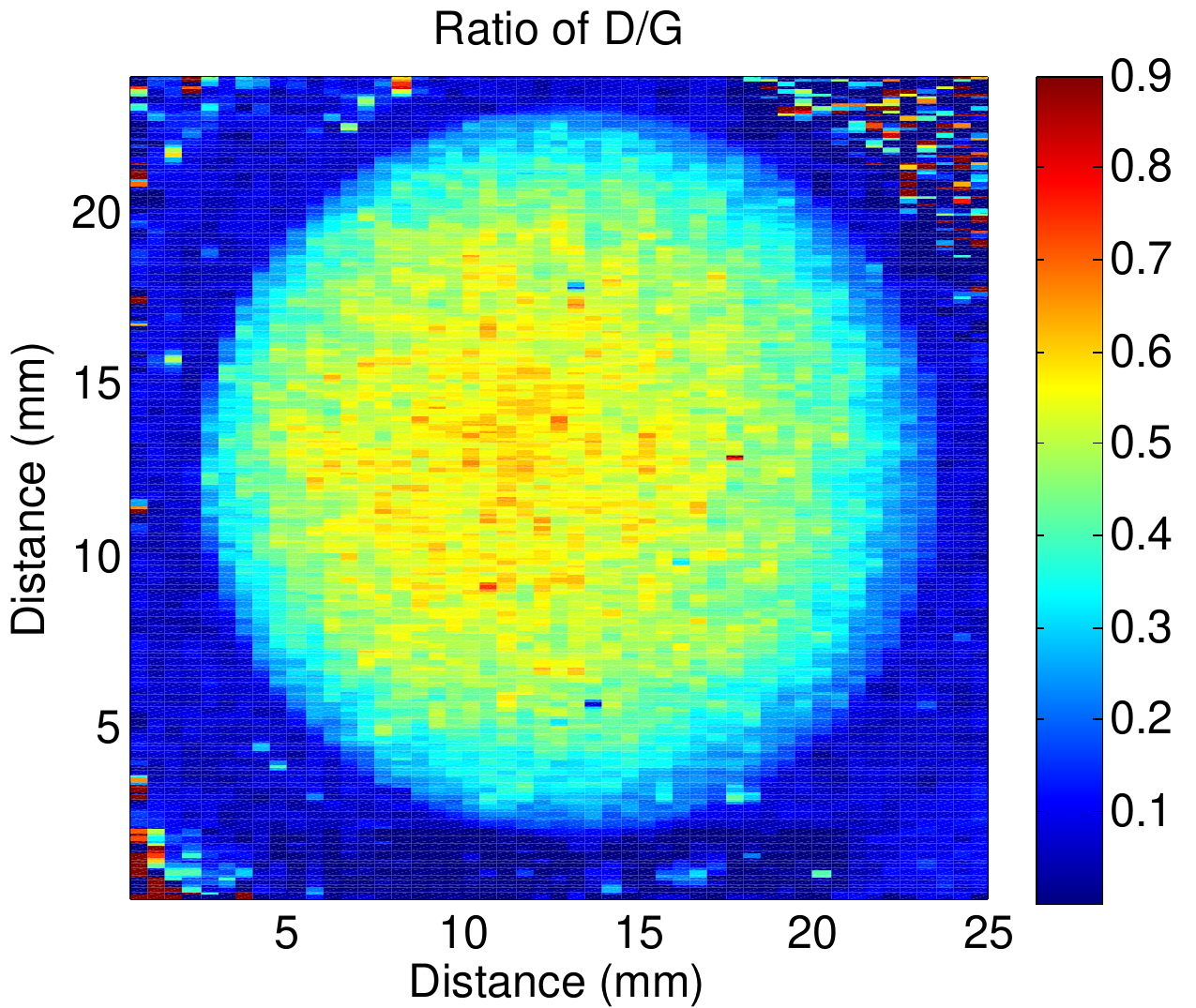}
                \caption{EUV+H$_2$}
                \label{fig:EUVH}
        \end{subfigure}~
        \begin{subfigure}{0.25\textwidth}
                \centering
                \includegraphics[width=\textwidth]{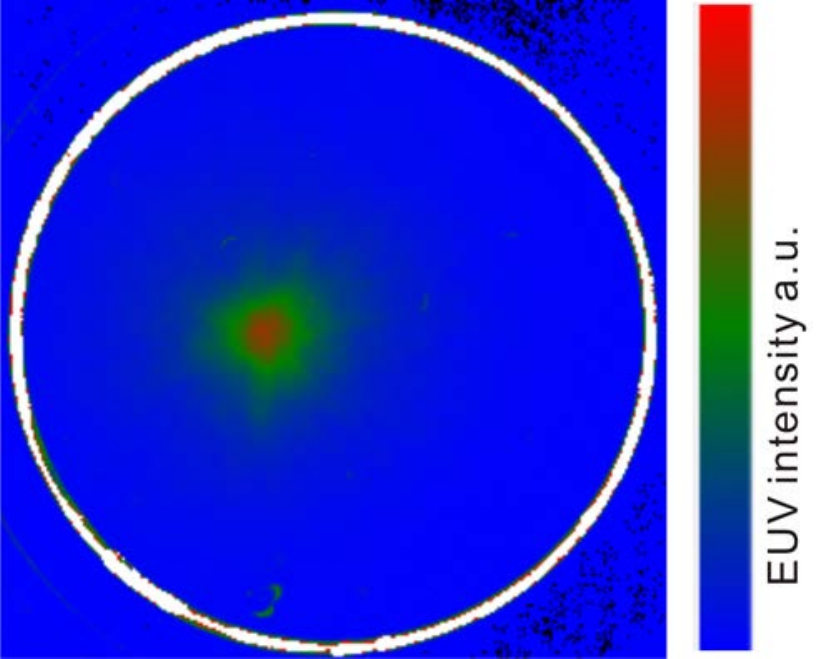}
                \caption{EUV Intensity profile}
               \label{fig:euvhprof}                
        \end{subfigure}
        \caption{(color online) \textit{I(D)/I(G)} ratio mapping. (b) and (c) are \textit{I(D)/I(G)} ratio maps for $S_{EUV}$ and $S_{EUV+H_2}$. (a) and (d) are the EUV intensity profiles for $S_{EUV}$ and $S_{EUV+H_2}$ respectively. The white circle indicates the mask boundary.}\label{fig:mappingb}
\end{figure*}

\begin{figure*}[!htb]
\begin{subfigure}{0.4\textwidth}
\centering
\includegraphics[width=\textwidth]{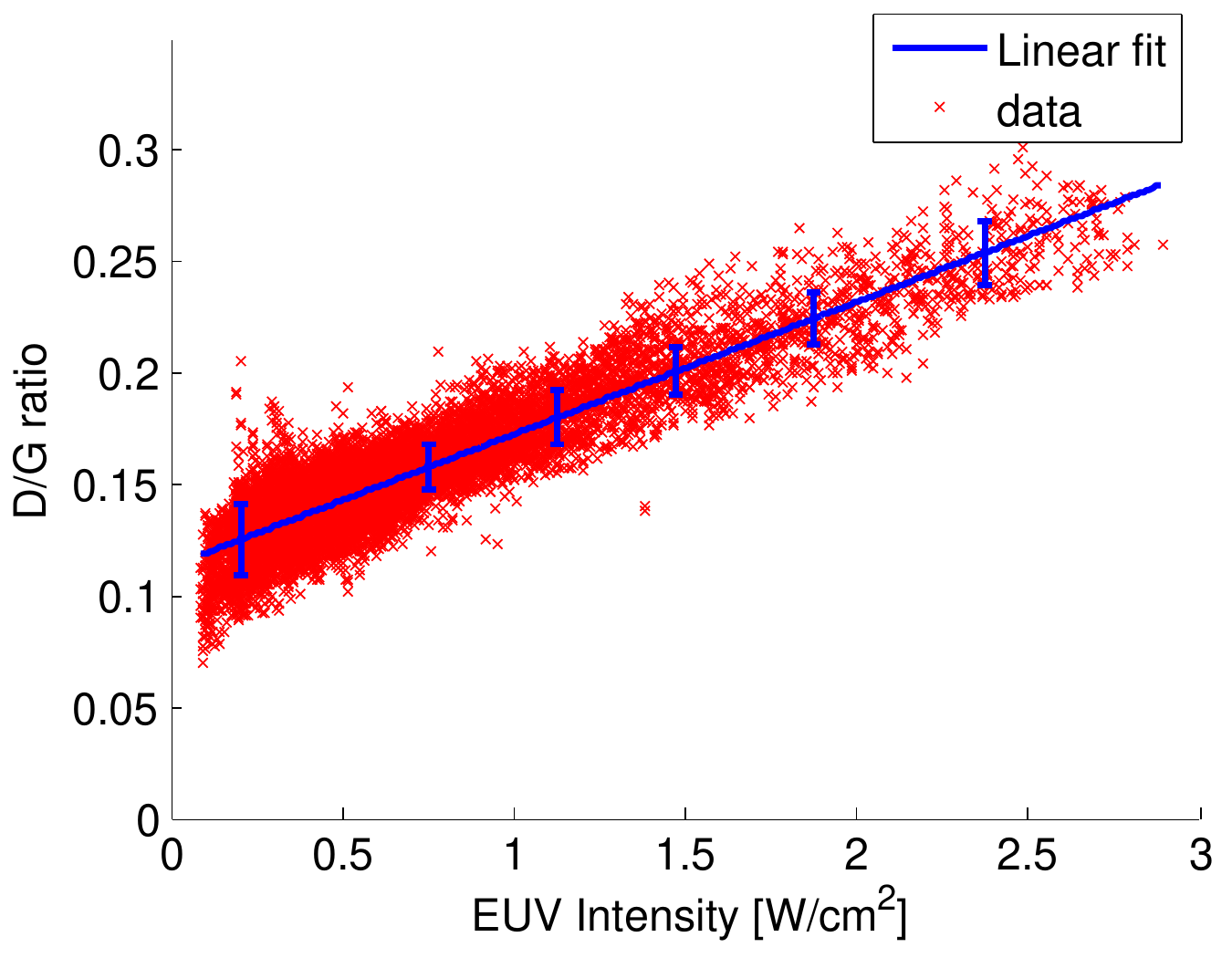}
\caption{EUV}
\label{fig:dose_euv}
\end{subfigure}~
\begin{subfigure}{0.4\textwidth}
\centering
\includegraphics[width=\textwidth]{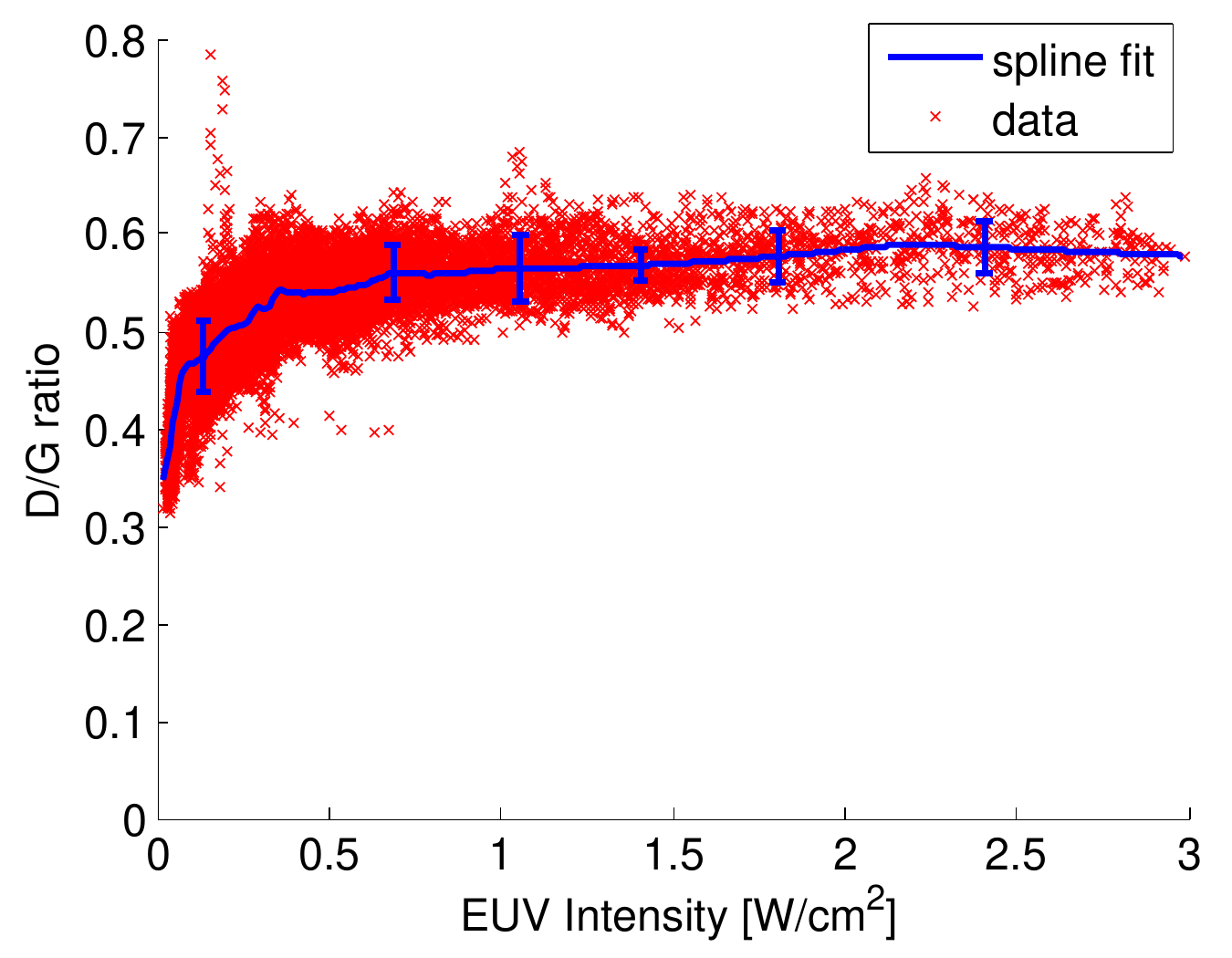}
\caption{EUV+H$_2$}
\label{fig:dose_euvh}
\end{subfigure}
\caption{(color online) \textit{I(D)/I(G)} ratio versus EUV power}\label{fig:dose_corr}
\end{figure*}

A typical Raman spectrum of graphene has three prominent features i.e., D, G and 2D peaks, located at 1350~c$m^{-1}$, 1580~$cm^{-1}$, and 2700~$cm^{-1}$ respectively. The G peak is a first order Raman scattering process, corresponding to an in plane streching of sp$^2$ bonds. The D band is due to the breathing modes of six-atom rings, and requires a defect for activation. The 2D peak is the second order of the D peak. Since the 2D originates from a process where momentum conservation is satisfied by two phonons with opposite wavevectors, defects are not required for their activation, and are, thus, always present~\cite{ferrari2006raman,ferrari2007raman}. Fig.~\ref{fig:spectra} shows the Raman spectra of the three samples. There is a small D peak in the spectrum of pristine sample, which is caused by natural defects such as edges, grain boundaries or vacancies in graphene produced by CVD~\cite{eckmann2012probing,gass2008free}. The spectrum for the sample that was exposed to EUV irradiation shows slightly higher D peak intensities compared to the pristine sample. The energetic photons from EUV irradiation might be expected to break sp$^2$ carbon bonds, leading to defects in graphene as well. The spectrum for the sample exposed to EUV in a hydrogen background has the highest D peak intensity. Besides the direct impact from EUV photons, hydrogen is photo-ionized by the EUV radiation, resulting in atomic and molecular hydrogen ions, atomic hydrogen, and electrons~\cite{chung1993dissociative,kossmann1989unexpected}. Energetic electrons are known to break carbon bonds forming defects in graphene~\cite{teweldebrhan2009modification,iqbal2012effect}. Furthermore, graphene hydrogenation occurs due to presence of a hydrogen plasma~\cite{elias2009control}. These combined effects lead to a higher defect density on the sample exposed to EUV in a hydrogen background. There is also a G peak shift from 1583~cm$^{-1}$ for pristine sample to 1598~cm$^{-1}$ for both $S_{EUV}$ and $S_{EUV+H_2}$, indicating the formation of sp$^2$ clusters or chains~\cite{ferrari2000interpretation,kudin2008raman}. Furthermore, there is another possible source for defects generation: secondary electrons from the Ni substrate, produced during EUV radiation. These electrons can be expected to have an energy less than 50~eV with a peak distribution between 2 and 5~eV~\cite{scholtz1996secondary}. These low energy electrons are not expected to create vacancy type defects. However, low energy electrons (7~eV) have been reported to dissociate adsorbed water and initiate oxide formation on metal surfaces~\cite{ebinger1998electron}. This remains to be investigated.      


\begin{figure*}[htb]
				\begin{subfigure}{0.45\textwidth}
                \centering
                \includegraphics[width=\textwidth]{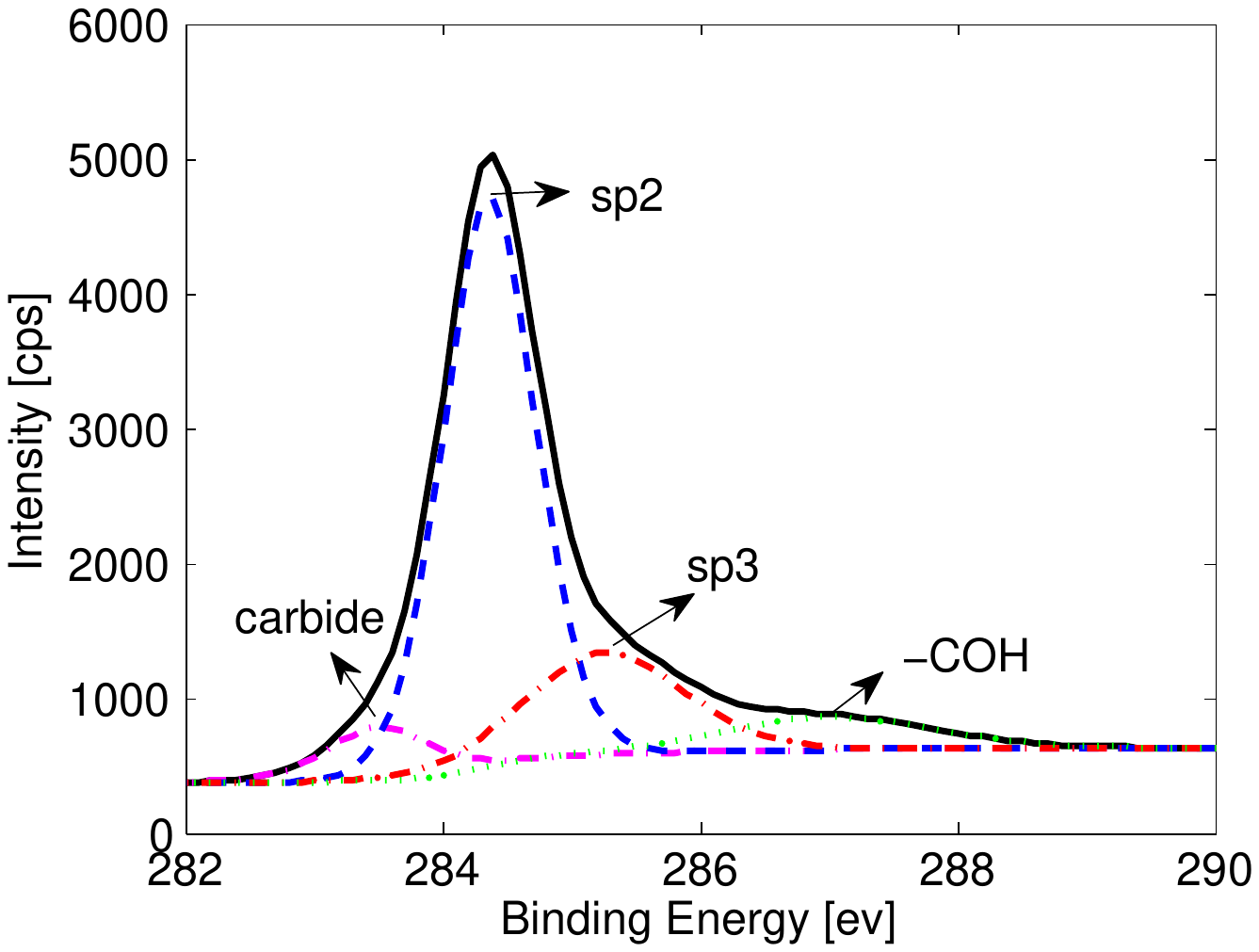}
                \caption{}
                \label{fig:euvhcf}
        \end{subfigure}~
        \begin{subfigure}{0.45\textwidth}
                \centering
                \includegraphics[width=\textwidth]{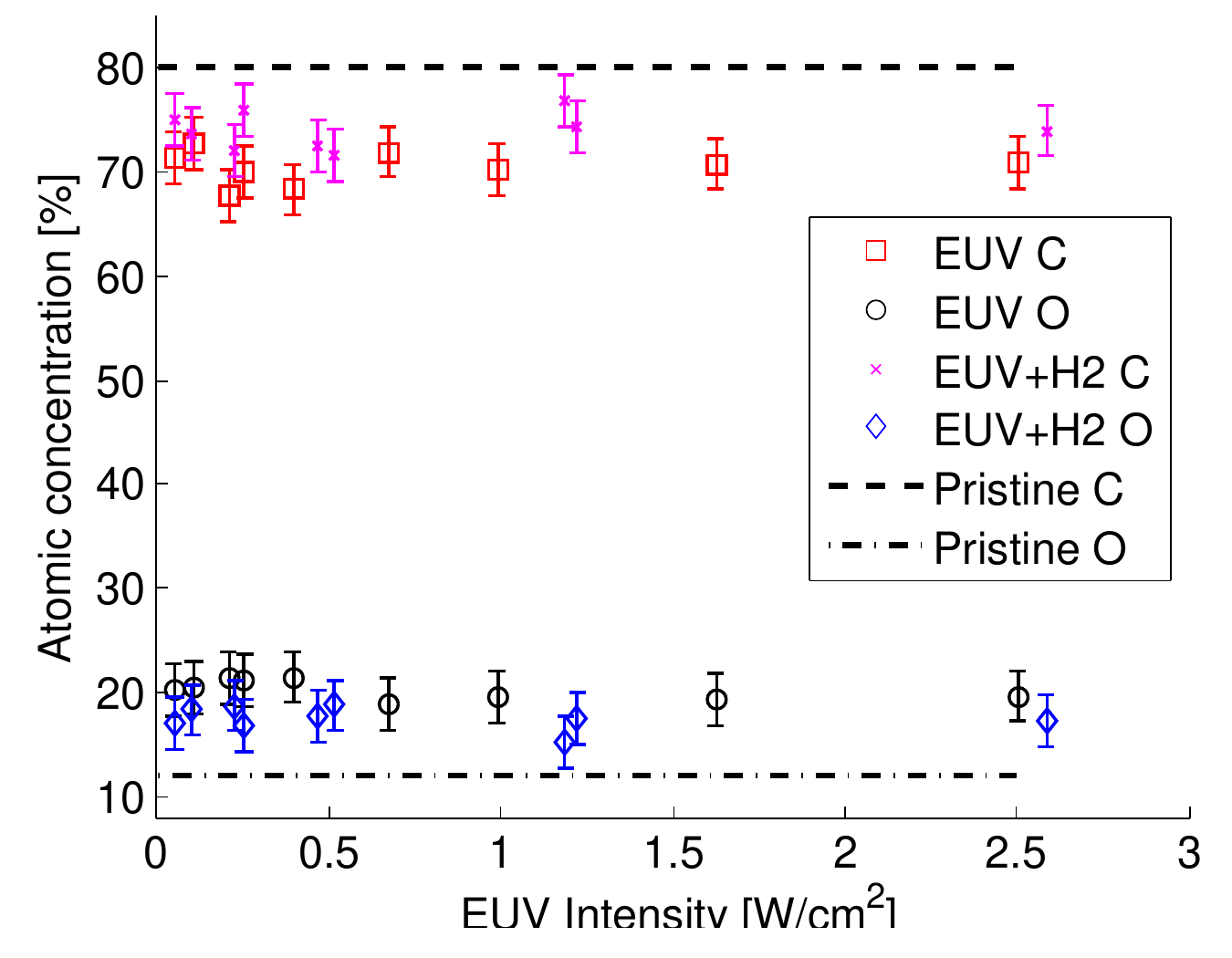}
                \caption{}
                \label{fig:element}
        \end{subfigure}\\
				\begin{subfigure}{0.45\textwidth}
                \centering
                \includegraphics[width=\textwidth]{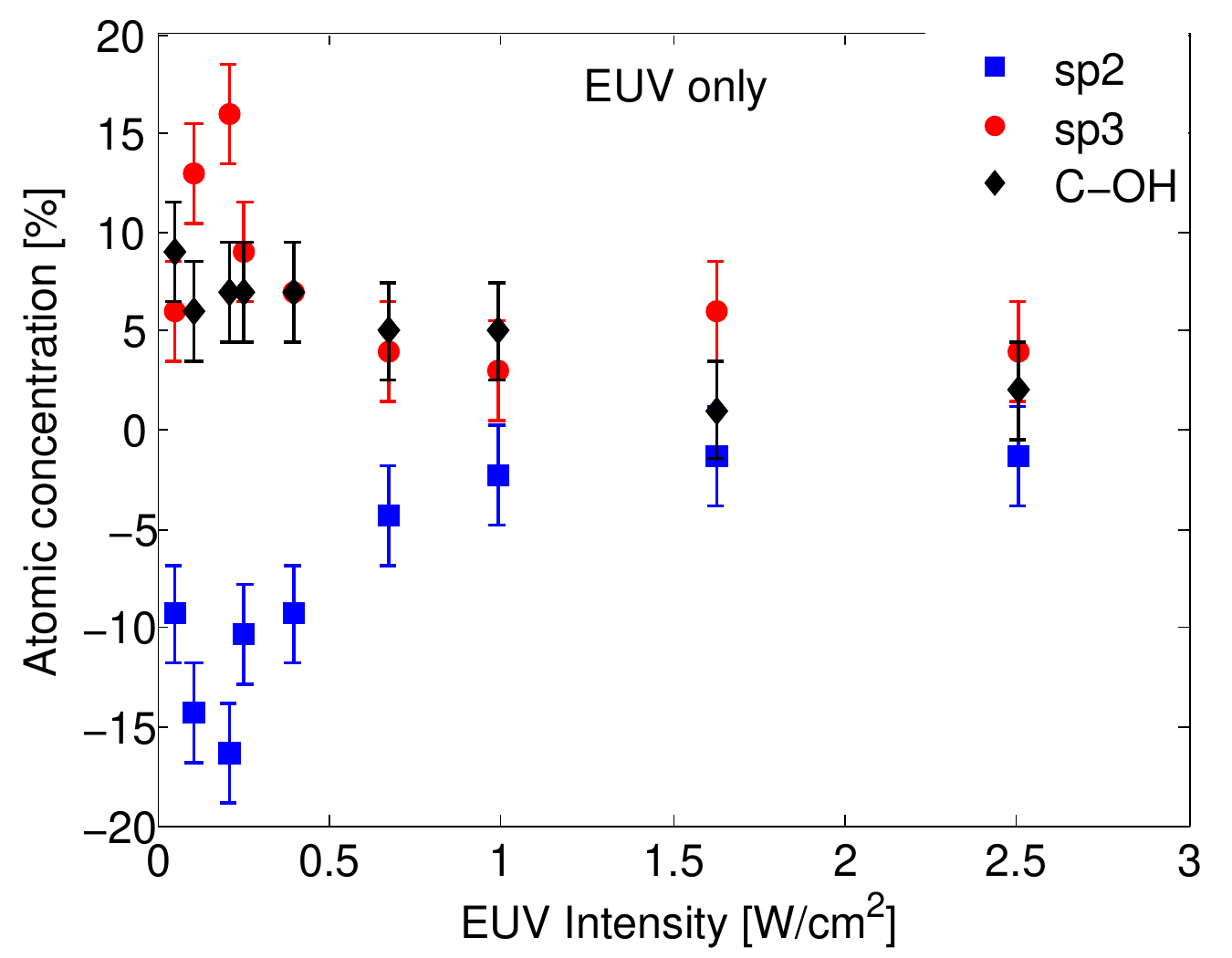}
                \caption{}
                \label{fig:euvc}
        \end{subfigure}~
        \begin{subfigure}{0.45\textwidth}
                \centering
                \includegraphics[width=\textwidth]{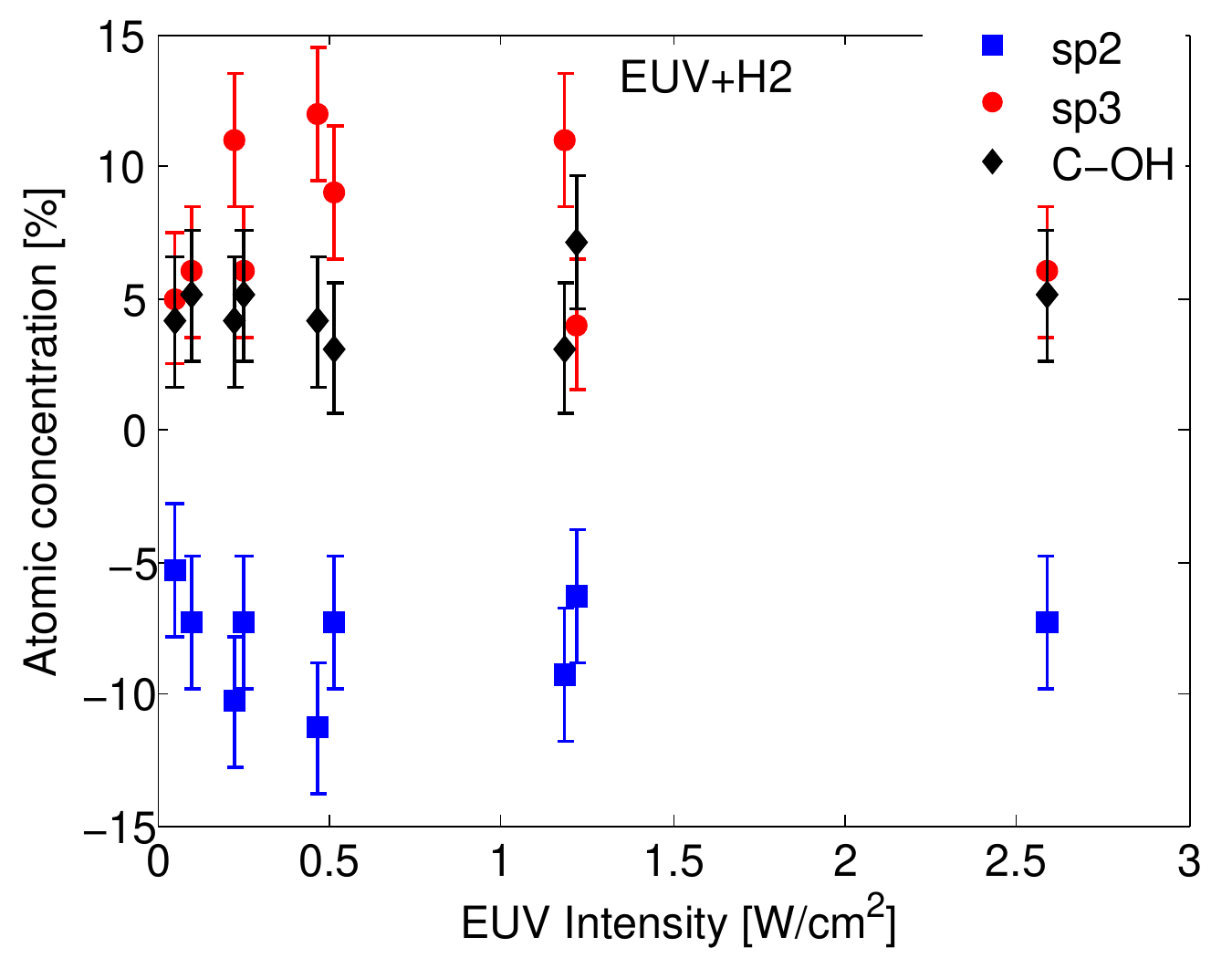}
                \caption{}
                \label{fig:euvhc}
        \end{subfigure}\\        
        
        \caption{(color online) (a) XPS analysis: curve fitting results for $S_{EUV}$; (b) Element concentration versus EUV power for $S_{EUV+H_2}$ and $S_{EUV}$; (c) and (d) Bond concentration change with respect to the pristine sample versus EUV power for $S_{EUV}$ and $S_{EUV+H_2}$.}\label{fig:xps}
\end{figure*}

Besides the single spectrum comparison, 2D scans for the two samples $S_{EUV}$ and $S_{EUV+H_2}$ were made to map the ratio of the D and G integrated intensities (shown in Fig.~\ref{fig:mappingb}). In Fig.~\ref{fig:mappingb}, the two samples $S_{EUV}$ and $S_{EUV+H_2}$ were partially covered with a metal mask. The spatial intensity distribution of EUV light is indicated in Fig.~\ref{fig:euvprof} and Fig.~\ref{fig:euvhprof}. Fig.~\ref{fig:EUVH} shows that $S_{EUV+H_2}$ has a higher D/G value, within exposed area, than that for $S_{EUV}$. It is also noted that for D/G ratio maps of the samples $S_{EUV}$ and $S_{EUV+H_2}$, there is a clear distinction between the exposed and unexposed areas. The D/G ratio map in Fig.~\ref{fig:EUV} clearly coincides with the EUV intensity profile shown in Fig.~\ref{fig:euvprof}. The D/G ratio is also plotted as a function of EUV dose for both $S_{EUV}$ and $S_{EUV+H_2}$ samples in Fig.~\ref{fig:dose_euv} and Fig.~\ref{fig:dose_euvh}. The D/G ratio first grows as the EUV intensity increases, then saturates. It appears that for $S_{EUV}$ the D/G ratio does not saturate as the EUV dose increases. Note that the I(D)/I(G) of $S_{EUV}$ value is lower than the ratio of $S_{EUV+H_2}$ , indicating that it may saturate at higher values. 

\subsection{XPS analysis}
Quantitative information on the relative concentrations of different C bond types in the sample were obtained by analyzing the C1s peak of the XPS spectrum~\cite{kaciulisspectroscopy,patsalas2001complementary}. The curve fitting results for the C1s spectrum of $S_{EUV+H_2}$ are shown in Fig.~\ref{fig:euvhcf}. There are four components in the C1s spectrum: the first peak at binding energy 283.4~eV, which is attributed to carbide formation with the underlying Ni layer,  the second peak at binding energy 284.4~eV, corresponds to the sp$^2$ bonds in graphitic like carbon, the third peak, at binding energy 285.3~eV, corresponds to carbon bonds with sp$^3$ hybridization, and the fourth peak, at binding energy 286.8~eV, is assigned to hydroxyl group.  The appearance of sp$^3$ carbon and C-OH both indicate the generation of defects in graphene. Oxidation occurs when graphene reacts with the residual water during exposure. At the same time, oxidation will generate at least one sp$^3$ bond as well. The sp$^3$ bonds can also be introduced by hydrogen plasma generated under EUV irradiation. In Fig.~\ref{fig:element}, for both the $S_{EUV+H_{2}}$ and $S_{EUV}$ sample, we can see that C element (the sp$^2$ bonded carbon) concentration drops by 5-9\% and O element concentration increases by 5-8\% compared with that in pristine sample.  The concentration change of different bonds versus EUV power with respect to the pristine sample are plotted in Fig.~\ref{fig:euvc} and Fig.~\ref{fig:euvhc}. In the case of $S_{EUV}$, the sp$^2$ concentration decreases less in the higher power range than in the lower power ranges. It appears that under EUV irradiation, besides breaking sp$^2$ bonds and forming sp$^3$ and C-OH bonds, there is also a transformation from C-OH phase to sp$^2$ phase, since the C-OH concentration change drops to almost zero. This transformation can be induced by local heating~\cite{bagri2010structural} due to EUV irradition. However, this transformation does not indicate that the converted sp$^2$ bonds are forming an ordered ring structure like in the undistorted graphene network, since, in the Raman spectrum, I(D)/I(G) (Fig.~\ref{fig:dose_euv}) increases in higher EUV power range.  In contrast, for $S_{EUV+H_{2}}$, the transformation to sp$^2$ is neglectable. Because hydrogenation can be the dominant effect, the converted sp$^2$ bonds will be hydrogenated in the end. Besides forming C-OH (oxidation), forming C-H bond (hydrogenation) will generate C-C (sp$^3$) bonds as well. The sp$^3$ concentration increases slowly at low intensities (lower than 0.5W/cm2) and saturates at higher powers, which coincide with the I(D)/I(G) ratio map in Fig.~\ref{fig:dose_euvh}. However, comparing $S_{EUV+H_{2}}$ with $S_{EUV}$, even with the same amount of sp$^2$, sp$^3$, and C-OH, they show different I(D)/I(G) values, indicating that there is no unique quantitative relationship between I(D)/I(G) ratio and sp$^3$ or C-OH content. The contribution from C-H or C-OH solely to I(D)/I(G) has yet to be investigated. Nevertheless, the XPS data clearly show that the defects were generated by EUV photons, including hydrogenation, and oxidation even in a reducing enviroment (H$_2$). 

\section{\label{sec:sum} Conclusion}The Raman results reported here show that there are defects induced in graphene after EUV irradiation, which is reflected by an increase of the D peak intensity. The defects are caused by breaking sp$^2$ bonds by EUV photons, oxidation due to the formation of OH groups, hydrogenation due to hydrogen plasma generated during EUV irradiation. The XPS results confirm that, after EUV irradiation, the concentration of sp$^2$ bonds in graphene decreases while the concentration of sp$^3$ bonds and C-OH bonds increases, clearly indicating defects generated in graphene. EUV irradiation introduces defects both through oxidation with the residual water background, and more effectively by hydrogenation due to the presence of hydrogen plasma. 

\begin{acknowledgments}
The authors would like to thank Mr. Goran Milinkovic, Mr. Luc Stevens, Mr. John de Kuster, and Dr. Edgar Osorio for the help with sample preparation and experimental measurements. This work is part of the research programme Controlling photon and plasma induced processes at EUV optical surfaces (CP3E) of the Stichting voor Fundamenteel Onderzoek der Materie (FOM) with financial support from the Nederlandse Organisatie voor Wetenschappelijk Onderzoek (NWO). The CP3E programme is co-financed by Carl Zeiss SMT and ASML, and the AgentschapNL through the EXEPT programme.
\end{acknowledgments}

\bibliography{euvref}

\begin{thebibliography}{30}%
\makeatletter
\providecommand \@ifxundefined [1]{%
 \@ifx{#1\undefined}
}%
\providecommand \@ifnum [1]{%
 \ifnum #1\expandafter \@firstoftwo
 \else \expandafter \@secondoftwo
 \fi
}%
\providecommand \@ifx [1]{%
 \ifx #1\expandafter \@firstoftwo
 \else \expandafter \@secondoftwo
 \fi
}%
\providecommand \natexlab [1]{#1}%
\providecommand \enquote  [1]{``#1''}%
\providecommand \bibnamefont  [1]{#1}%
\providecommand \bibfnamefont [1]{#1}%
\providecommand \citenamefont [1]{#1}%
\providecommand \href@noop [0]{\@secondoftwo}%
\providecommand \href [0]{\begingroup \@sanitize@url \@href}%
\providecommand \@href[1]{\@@startlink{#1}\@@href}%
\providecommand \@@href[1]{\endgroup#1\@@endlink}%
\providecommand \@sanitize@url [0]{\catcode `\\12\catcode `\$12\catcode
  `\&12\catcode `\#12\catcode `\^12\catcode `\_12\catcode `\%12\relax}%
\providecommand \@@startlink[1]{}%
\providecommand \@@endlink[0]{}%
\providecommand \url  [0]{\begingroup\@sanitize@url \@url }%
\providecommand \@url [1]{\endgroup\@href {#1}{\urlprefix }}%
\providecommand \urlprefix  [0]{URL }%
\providecommand \Eprint [0]{\href }%
\providecommand \doibase [0]{http://dx.doi.org/}%
\providecommand \selectlanguage [0]{\@gobble}%
\providecommand \bibinfo  [0]{\@secondoftwo}%
\providecommand \bibfield  [0]{\@secondoftwo}%
\providecommand \translation [1]{[#1]}%
\providecommand \BibitemOpen [0]{}%
\providecommand \bibitemStop [0]{}%
\providecommand \bibitemNoStop [0]{.\EOS\space}%
\providecommand \EOS [0]{\spacefactor3000\relax}%
\providecommand \BibitemShut  [1]{\csname bibitem#1\endcsname}%
\let\auto@bib@innerbib\@empty
\bibitem [{\citenamefont {Geim}\ and\ \citenamefont
  {Novoselov}(2007)}]{geim2007rise}%
  \BibitemOpen
  \bibfield  {author} {\bibinfo {author} {\bibfnamefont {A.}~\bibnamefont
  {Geim}}\ and\ \bibinfo {author} {\bibfnamefont {K.}~\bibnamefont
  {Novoselov}},\ }\bibfield  {title} {\enquote {\bibinfo {title} {The rise of
  graphene},}\ }\href@noop {} {\bibfield  {journal} {\bibinfo  {journal}
  {Nature materials}\ }\textbf {\bibinfo {volume} {6}},\ \bibinfo {pages}
  {183--191} (\bibinfo {year} {2007})}\BibitemShut {NoStop}%
\bibitem [{\citenamefont {Geim}(2009)}]{geim2009graphene}%
  \BibitemOpen
  \bibfield  {author} {\bibinfo {author} {\bibfnamefont {A.}~\bibnamefont
  {Geim}},\ }\bibfield  {title} {\enquote {\bibinfo {title} {Graphene: status
  and prospects},}\ }\href@noop {} {\bibfield  {journal} {\bibinfo  {journal}
  {science}\ }\textbf {\bibinfo {volume} {324}},\ \bibinfo {pages} {1530--1534}
  (\bibinfo {year} {2009})}\BibitemShut {NoStop}%
\bibitem [{\citenamefont {Novoselov}\ \emph {et~al.}(2005)\citenamefont
  {Novoselov}, \citenamefont {Jiang}, \citenamefont {Schedin}, \citenamefont
  {Booth}, \citenamefont {Khotkevich}, \citenamefont {Morozov},\ and\
  \citenamefont {Geim}}]{novoselov2005two}%
  \BibitemOpen
  \bibfield  {author} {\bibinfo {author} {\bibfnamefont {K.}~\bibnamefont
  {Novoselov}}, \bibinfo {author} {\bibfnamefont {D.}~\bibnamefont {Jiang}},
  \bibinfo {author} {\bibfnamefont {F.}~\bibnamefont {Schedin}}, \bibinfo
  {author} {\bibfnamefont {T.}~\bibnamefont {Booth}}, \bibinfo {author}
  {\bibfnamefont {V.}~\bibnamefont {Khotkevich}}, \bibinfo {author}
  {\bibfnamefont {S.}~\bibnamefont {Morozov}}, \ and\ \bibinfo {author}
  {\bibfnamefont {A.}~\bibnamefont {Geim}},\ }\bibfield  {title} {\enquote
  {\bibinfo {title} {Two-dimensional atomic crystals},}\ }\href@noop {}
  {\bibfield  {journal} {\bibinfo  {journal} {Proceedings of the National
  Academy of Sciences of the United States of America}\ }\textbf {\bibinfo
  {volume} {102}},\ \bibinfo {pages} {10451} (\bibinfo {year}
  {2005})}\BibitemShut {NoStop}%
\bibitem [{\citenamefont {Zhang}\ \emph {et~al.}(2005)\citenamefont {Zhang},
  \citenamefont {Tan}, \citenamefont {Stormer},\ and\ \citenamefont
  {Kim}}]{zhang2005experimental}%
  \BibitemOpen
  \bibfield  {author} {\bibinfo {author} {\bibfnamefont {Y.}~\bibnamefont
  {Zhang}}, \bibinfo {author} {\bibfnamefont {Y.}~\bibnamefont {Tan}}, \bibinfo
  {author} {\bibfnamefont {H.}~\bibnamefont {Stormer}}, \ and\ \bibinfo
  {author} {\bibfnamefont {P.}~\bibnamefont {Kim}},\ }\bibfield  {title}
  {\enquote {\bibinfo {title} {Experimental observation of the quantum hall
  effect and berry's phase in graphene},}\ }\href@noop {} {\bibfield  {journal}
  {\bibinfo  {journal} {Nature}\ }\textbf {\bibinfo {volume} {438}},\ \bibinfo
  {pages} {201--204} (\bibinfo {year} {2005})}\BibitemShut {NoStop}%
\bibitem [{\citenamefont {Han}\ \emph {et~al.}(2007)\citenamefont {Han},
  \citenamefont {{\"O}zyilmaz}, \citenamefont {Zhang},\ and\ \citenamefont
  {Kim}}]{han2007energy}%
  \BibitemOpen
  \bibfield  {author} {\bibinfo {author} {\bibfnamefont {M.}~\bibnamefont
  {Han}}, \bibinfo {author} {\bibfnamefont {B.}~\bibnamefont {{\"O}zyilmaz}},
  \bibinfo {author} {\bibfnamefont {Y.}~\bibnamefont {Zhang}}, \ and\ \bibinfo
  {author} {\bibfnamefont {P.}~\bibnamefont {Kim}},\ }\bibfield  {title}
  {\enquote {\bibinfo {title} {Energy band-gap engineering of graphene
  nanoribbons},}\ }\href@noop {} {\bibfield  {journal} {\bibinfo  {journal}
  {Physical Review Letters}\ }\textbf {\bibinfo {volume} {98}},\ \bibinfo
  {pages} {206805} (\bibinfo {year} {2007})}\BibitemShut {NoStop}%
\bibitem [{\citenamefont {Bolotin}\ \emph {et~al.}(2008)\citenamefont
  {Bolotin}, \citenamefont {Sikes}, \citenamefont {Jiang}, \citenamefont
  {Klima}, \citenamefont {Fudenberg}, \citenamefont {Hone}, \citenamefont
  {Kim},\ and\ \citenamefont {Stormer}}]{bolotin2008ultrahigh}%
  \BibitemOpen
  \bibfield  {author} {\bibinfo {author} {\bibfnamefont {K.}~\bibnamefont
  {Bolotin}}, \bibinfo {author} {\bibfnamefont {K.}~\bibnamefont {Sikes}},
  \bibinfo {author} {\bibfnamefont {Z.}~\bibnamefont {Jiang}}, \bibinfo
  {author} {\bibfnamefont {M.}~\bibnamefont {Klima}}, \bibinfo {author}
  {\bibfnamefont {G.}~\bibnamefont {Fudenberg}}, \bibinfo {author}
  {\bibfnamefont {J.}~\bibnamefont {Hone}}, \bibinfo {author} {\bibfnamefont
  {P.}~\bibnamefont {Kim}}, \ and\ \bibinfo {author} {\bibfnamefont
  {H.}~\bibnamefont {Stormer}},\ }\bibfield  {title} {\enquote {\bibinfo
  {title} {Ultrahigh electron mobility in suspended graphene},}\ }\href@noop {}
  {\bibfield  {journal} {\bibinfo  {journal} {Solid State Communications}\
  }\textbf {\bibinfo {volume} {146}},\ \bibinfo {pages} {351--355} (\bibinfo
  {year} {2008})}\BibitemShut {NoStop}%
\bibitem [{\citenamefont {Lee}\ \emph {et~al.}(2008)\citenamefont {Lee},
  \citenamefont {Wei}, \citenamefont {Kysar},\ and\ \citenamefont
  {Hone}}]{lee2008measurement}%
  \BibitemOpen
  \bibfield  {author} {\bibinfo {author} {\bibfnamefont {C.}~\bibnamefont
  {Lee}}, \bibinfo {author} {\bibfnamefont {X.}~\bibnamefont {Wei}}, \bibinfo
  {author} {\bibfnamefont {J.}~\bibnamefont {Kysar}}, \ and\ \bibinfo {author}
  {\bibfnamefont {J.}~\bibnamefont {Hone}},\ }\bibfield  {title} {\enquote
  {\bibinfo {title} {Measurement of the elastic properties and intrinsic
  strength of monolayer graphene},}\ }\href@noop {} {\bibfield  {journal}
  {\bibinfo  {journal} {Science}\ }\textbf {\bibinfo {volume} {321}},\ \bibinfo
  {pages} {385--388} (\bibinfo {year} {2008})}\BibitemShut {NoStop}%
\bibitem [{\citenamefont {Bunch}\ \emph {et~al.}(2007)\citenamefont {Bunch},
  \citenamefont {Van~der Zande}, \citenamefont {Verbridge}, \citenamefont
  {Frank}, \citenamefont {Tanenbaum}, \citenamefont {Parpia}, \citenamefont
  {Craighead},\ and\ \citenamefont {McEuen}}]{bunch2007electromechanical}%
  \BibitemOpen
  \bibfield  {author} {\bibinfo {author} {\bibfnamefont {J.}~\bibnamefont
  {Bunch}}, \bibinfo {author} {\bibfnamefont {A.}~\bibnamefont {Van~der
  Zande}}, \bibinfo {author} {\bibfnamefont {S.}~\bibnamefont {Verbridge}},
  \bibinfo {author} {\bibfnamefont {I.}~\bibnamefont {Frank}}, \bibinfo
  {author} {\bibfnamefont {D.}~\bibnamefont {Tanenbaum}}, \bibinfo {author}
  {\bibfnamefont {J.}~\bibnamefont {Parpia}}, \bibinfo {author} {\bibfnamefont
  {H.}~\bibnamefont {Craighead}}, \ and\ \bibinfo {author} {\bibfnamefont
  {P.}~\bibnamefont {McEuen}},\ }\bibfield  {title} {\enquote {\bibinfo {title}
  {Electromechanical resonators from graphene sheets},}\ }\href@noop {}
  {\bibfield  {journal} {\bibinfo  {journal} {Science}\ }\textbf {\bibinfo
  {volume} {315}},\ \bibinfo {pages} {490--493} (\bibinfo {year}
  {2007})}\BibitemShut {NoStop}%
\bibitem [{\citenamefont {Boukhvalov}\ and\ \citenamefont
  {Katsnelson}(2008)}]{boukhvalov2008chemical}%
  \BibitemOpen
  \bibfield  {author} {\bibinfo {author} {\bibfnamefont {D.}~\bibnamefont
  {Boukhvalov}}\ and\ \bibinfo {author} {\bibfnamefont {M.}~\bibnamefont
  {Katsnelson}},\ }\bibfield  {title} {\enquote {\bibinfo {title} {Chemical
  functionalization of graphene with defects},}\ }\href@noop {} {\bibfield
  {journal} {\bibinfo  {journal} {Nano letters}\ }\textbf {\bibinfo {volume}
  {8}},\ \bibinfo {pages} {4373--4379} (\bibinfo {year} {2008})}\BibitemShut
  {NoStop}%
\bibitem [{\citenamefont {Cortijo}\ and\ \citenamefont
  {Vozmediano}(2007)}]{cortijo2007effects}%
  \BibitemOpen
  \bibfield  {author} {\bibinfo {author} {\bibfnamefont {A.}~\bibnamefont
  {Cortijo}}\ and\ \bibinfo {author} {\bibfnamefont {M.}~\bibnamefont
  {Vozmediano}},\ }\bibfield  {title} {\enquote {\bibinfo {title} {Effects of
  topological defects and local curvature on the electronic properties of
  planar graphene},}\ }\href@noop {} {\bibfield  {journal} {\bibinfo  {journal}
  {Nuclear Physics B}\ }\textbf {\bibinfo {volume} {763}},\ \bibinfo {pages}
  {293--308} (\bibinfo {year} {2007})}\BibitemShut {NoStop}%
\bibitem [{\citenamefont {Rutter}\ \emph {et~al.}(2007)\citenamefont {Rutter},
  \citenamefont {Crain}, \citenamefont {Guisinger}, \citenamefont {Li},
  \citenamefont {First},\ and\ \citenamefont
  {Stroscio}}]{rutter2007scattering}%
  \BibitemOpen
  \bibfield  {author} {\bibinfo {author} {\bibfnamefont {G.}~\bibnamefont
  {Rutter}}, \bibinfo {author} {\bibfnamefont {J.}~\bibnamefont {Crain}},
  \bibinfo {author} {\bibfnamefont {N.}~\bibnamefont {Guisinger}}, \bibinfo
  {author} {\bibfnamefont {T.}~\bibnamefont {Li}}, \bibinfo {author}
  {\bibfnamefont {P.}~\bibnamefont {First}}, \ and\ \bibinfo {author}
  {\bibfnamefont {J.}~\bibnamefont {Stroscio}},\ }\bibfield  {title} {\enquote
  {\bibinfo {title} {Scattering and interference in epitaxial graphene},}\
  }\href@noop {} {\bibfield  {journal} {\bibinfo  {journal} {Science}\ }\textbf
  {\bibinfo {volume} {317}},\ \bibinfo {pages} {219--222} (\bibinfo {year}
  {2007})}\BibitemShut {NoStop}%
\bibitem [{\citenamefont {Hao}, \citenamefont {Fang},\ and\ \citenamefont
  {Xu}(2011)}]{hao2011mechanical}%
  \BibitemOpen
  \bibfield  {author} {\bibinfo {author} {\bibfnamefont {F.}~\bibnamefont
  {Hao}}, \bibinfo {author} {\bibfnamefont {D.}~\bibnamefont {Fang}}, \ and\
  \bibinfo {author} {\bibfnamefont {Z.}~\bibnamefont {Xu}},\ }\bibfield
  {title} {\enquote {\bibinfo {title} {Mechanical and thermal transport
  properties of graphene with defects},}\ }\href@noop {} {\bibfield  {journal}
  {\bibinfo  {journal} {Applied Physics Letters}\ }\textbf {\bibinfo {volume}
  {99}},\ \bibinfo {pages} {041901--041901} (\bibinfo {year}
  {2011})}\BibitemShut {NoStop}%
\bibitem [{\citenamefont {Zhou}\ \emph {et~al.}(2009)\citenamefont {Zhou},
  \citenamefont {Girit}, \citenamefont {Scholl}, \citenamefont {Jozwiak},
  \citenamefont {Siegel}, \citenamefont {Yu}, \citenamefont {Robinson},
  \citenamefont {Wang}, \citenamefont {Zettl},\ and\ \citenamefont
  {Lanzara}}]{zhou2009instability}%
  \BibitemOpen
  \bibfield  {author} {\bibinfo {author} {\bibfnamefont {S.}~\bibnamefont
  {Zhou}}, \bibinfo {author} {\bibfnamefont {{\c{C}}.}~\bibnamefont {Girit}},
  \bibinfo {author} {\bibfnamefont {A.}~\bibnamefont {Scholl}}, \bibinfo
  {author} {\bibfnamefont {C.}~\bibnamefont {Jozwiak}}, \bibinfo {author}
  {\bibfnamefont {D.}~\bibnamefont {Siegel}}, \bibinfo {author} {\bibfnamefont
  {P.}~\bibnamefont {Yu}}, \bibinfo {author} {\bibfnamefont {J.}~\bibnamefont
  {Robinson}}, \bibinfo {author} {\bibfnamefont {F.}~\bibnamefont {Wang}},
  \bibinfo {author} {\bibfnamefont {A.}~\bibnamefont {Zettl}}, \ and\ \bibinfo
  {author} {\bibfnamefont {A.}~\bibnamefont {Lanzara}},\ }\bibfield  {title}
  {\enquote {\bibinfo {title} {Instability of two-dimensional graphene:
  Breaking sp$^2$ bonds with soft x rays},}\ }\href@noop {} {\bibfield
  {journal} {\bibinfo  {journal} {Physical Review B}\ }\textbf {\bibinfo
  {volume} {80}},\ \bibinfo {pages} {121409} (\bibinfo {year}
  {2009})}\BibitemShut {NoStop}%
\bibitem [{\citenamefont {Hicks}\ \emph {et~al.}(2011)\citenamefont {Hicks},
  \citenamefont {Arora}, \citenamefont {Kenyon}, \citenamefont {Chakraborty},
  \citenamefont {Tinkey}, \citenamefont {Hankinson}, \citenamefont {Berger},
  \citenamefont {de~Heer}, \citenamefont {Conrad},\ and\ \citenamefont
  {Cressler}}]{hicks2011x}%
  \BibitemOpen
  \bibfield  {author} {\bibinfo {author} {\bibfnamefont {J.}~\bibnamefont
  {Hicks}}, \bibinfo {author} {\bibfnamefont {R.}~\bibnamefont {Arora}},
  \bibinfo {author} {\bibfnamefont {E.}~\bibnamefont {Kenyon}}, \bibinfo
  {author} {\bibfnamefont {P.}~\bibnamefont {Chakraborty}}, \bibinfo {author}
  {\bibfnamefont {H.}~\bibnamefont {Tinkey}}, \bibinfo {author} {\bibfnamefont
  {J.}~\bibnamefont {Hankinson}}, \bibinfo {author} {\bibfnamefont
  {C.}~\bibnamefont {Berger}}, \bibinfo {author} {\bibfnamefont
  {W.}~\bibnamefont {de~Heer}}, \bibinfo {author} {\bibfnamefont
  {E.}~\bibnamefont {Conrad}}, \ and\ \bibinfo {author} {\bibfnamefont
  {J.}~\bibnamefont {Cressler}},\ }\bibfield  {title} {\enquote {\bibinfo
  {title} {X-ray radiation effects in multilayer epitaxial graphene},}\
  }\href@noop {} {\bibfield  {journal} {\bibinfo  {journal} {Applied Physics
  Letters}\ }\textbf {\bibinfo {volume} {99}},\ \bibinfo {pages}
  {232102--232102} (\bibinfo {year} {2011})}\BibitemShut {NoStop}%
\bibitem [{\citenamefont {Ferrari}\ \emph {et~al.}(2006)\citenamefont
  {Ferrari}, \citenamefont {Meyer}, \citenamefont {Scardaci}, \citenamefont
  {Casiraghi}, \citenamefont {Lazzeri}, \citenamefont {Mauri}, \citenamefont
  {Piscanec}, \citenamefont {Jiang}, \citenamefont {Novoselov}, \citenamefont
  {Roth} \emph {et~al.}}]{ferrari2006raman}%
  \BibitemOpen
  \bibfield  {author} {\bibinfo {author} {\bibfnamefont {A.}~\bibnamefont
  {Ferrari}}, \bibinfo {author} {\bibfnamefont {J.}~\bibnamefont {Meyer}},
  \bibinfo {author} {\bibfnamefont {V.}~\bibnamefont {Scardaci}}, \bibinfo
  {author} {\bibfnamefont {C.}~\bibnamefont {Casiraghi}}, \bibinfo {author}
  {\bibfnamefont {M.}~\bibnamefont {Lazzeri}}, \bibinfo {author} {\bibfnamefont
  {F.}~\bibnamefont {Mauri}}, \bibinfo {author} {\bibfnamefont
  {S.}~\bibnamefont {Piscanec}}, \bibinfo {author} {\bibfnamefont
  {D.}~\bibnamefont {Jiang}}, \bibinfo {author} {\bibfnamefont
  {K.}~\bibnamefont {Novoselov}}, \bibinfo {author} {\bibfnamefont
  {S.}~\bibnamefont {Roth}},  \emph {et~al.},\ }\bibfield  {title} {\enquote
  {\bibinfo {title} {Raman spectrum of graphene and graphene layers},}\
  }\href@noop {} {\bibfield  {journal} {\bibinfo  {journal} {Physical Review
  Letters}\ }\textbf {\bibinfo {volume} {97}},\ \bibinfo {pages} {187401}
  (\bibinfo {year} {2006})}\BibitemShut {NoStop}%
\bibitem [{\citenamefont {Ferrari}(2007)}]{ferrari2007raman}%
  \BibitemOpen
  \bibfield  {author} {\bibinfo {author} {\bibfnamefont {A.}~\bibnamefont
  {Ferrari}},\ }\bibfield  {title} {\enquote {\bibinfo {title} {Raman
  spectroscopy of graphene and graphite: Disorder, electron--phonon coupling,
  doping and nonadiabatic effects},}\ }\href@noop {} {\bibfield  {journal}
  {\bibinfo  {journal} {Solid State Communications}\ }\textbf {\bibinfo
  {volume} {143}},\ \bibinfo {pages} {47--57} (\bibinfo {year}
  {2007})}\BibitemShut {NoStop}%
\bibitem [{\citenamefont {Eckmann}\ \emph {et~al.}(2012)\citenamefont
  {Eckmann}, \citenamefont {Felten}, \citenamefont {Mishchenko}, \citenamefont
  {Britnell}, \citenamefont {Krupke}, \citenamefont {Novoselov},\ and\
  \citenamefont {Casiraghi}}]{eckmann2012probing}%
  \BibitemOpen
  \bibfield  {author} {\bibinfo {author} {\bibfnamefont {A.}~\bibnamefont
  {Eckmann}}, \bibinfo {author} {\bibfnamefont {A.}~\bibnamefont {Felten}},
  \bibinfo {author} {\bibfnamefont {A.}~\bibnamefont {Mishchenko}}, \bibinfo
  {author} {\bibfnamefont {L.}~\bibnamefont {Britnell}}, \bibinfo {author}
  {\bibfnamefont {R.}~\bibnamefont {Krupke}}, \bibinfo {author} {\bibfnamefont
  {K.}~\bibnamefont {Novoselov}}, \ and\ \bibinfo {author} {\bibfnamefont
  {C.}~\bibnamefont {Casiraghi}},\ }\bibfield  {title} {\enquote {\bibinfo
  {title} {Probing the nature of defects in graphene by raman spectroscopy},}\
  }\href@noop {} {\bibfield  {journal} {\bibinfo  {journal} {Nano Letters}\ }
  (\bibinfo {year} {2012})}\BibitemShut {NoStop}%
\bibitem [{\citenamefont {Gass}\ \emph {et~al.}(2008)\citenamefont {Gass},
  \citenamefont {Bangert}, \citenamefont {Bleloch}, \citenamefont {Wang},
  \citenamefont {Nair},\ and\ \citenamefont {Geim}}]{gass2008free}%
  \BibitemOpen
  \bibfield  {author} {\bibinfo {author} {\bibfnamefont {M.}~\bibnamefont
  {Gass}}, \bibinfo {author} {\bibfnamefont {U.}~\bibnamefont {Bangert}},
  \bibinfo {author} {\bibfnamefont {A.}~\bibnamefont {Bleloch}}, \bibinfo
  {author} {\bibfnamefont {P.}~\bibnamefont {Wang}}, \bibinfo {author}
  {\bibfnamefont {R.}~\bibnamefont {Nair}}, \ and\ \bibinfo {author}
  {\bibfnamefont {A.}~\bibnamefont {Geim}},\ }\bibfield  {title} {\enquote
  {\bibinfo {title} {Free-standing graphene at atomic resolution},}\
  }\href@noop {} {\bibfield  {journal} {\bibinfo  {journal} {Nature
  nanotechnology}\ }\textbf {\bibinfo {volume} {3}},\ \bibinfo {pages}
  {676--681} (\bibinfo {year} {2008})}\BibitemShut {NoStop}%
\bibitem [{\citenamefont {Chung}\ \emph {et~al.}(1993)\citenamefont {Chung},
  \citenamefont {Lee}, \citenamefont {Masuoka},\ and\ \citenamefont
  {Samson}}]{chung1993dissociative}%
  \BibitemOpen
  \bibfield  {author} {\bibinfo {author} {\bibfnamefont {Y.}~\bibnamefont
  {Chung}}, \bibinfo {author} {\bibfnamefont {E.}~\bibnamefont {Lee}}, \bibinfo
  {author} {\bibfnamefont {T.}~\bibnamefont {Masuoka}}, \ and\ \bibinfo
  {author} {\bibfnamefont {J.}~\bibnamefont {Samson}},\ }\bibfield  {title}
  {\enquote {\bibinfo {title} {Dissociative photoionization of h from 18 to 124
  ev},}\ }\href@noop {} {\bibfield  {journal} {\bibinfo  {journal} {The Journal
  of chemical physics}\ }\textbf {\bibinfo {volume} {99}},\ \bibinfo {pages}
  {885} (\bibinfo {year} {1993})}\BibitemShut {NoStop}%
\bibitem [{\citenamefont {Kossmann}\ \emph {et~al.}(1989)\citenamefont
  {Kossmann}, \citenamefont {Schwarzkopf}, \citenamefont {K{\"a}mmerling},\
  and\ \citenamefont {Schmidt}}]{kossmann1989unexpected}%
  \BibitemOpen
  \bibfield  {author} {\bibinfo {author} {\bibfnamefont {H.}~\bibnamefont
  {Kossmann}}, \bibinfo {author} {\bibfnamefont {O.}~\bibnamefont
  {Schwarzkopf}}, \bibinfo {author} {\bibfnamefont {B.}~\bibnamefont
  {K{\"a}mmerling}}, \ and\ \bibinfo {author} {\bibfnamefont {V.}~\bibnamefont
  {Schmidt}},\ }\bibfield  {title} {\enquote {\bibinfo {title} {Unexpected
  behaviour of double photoionization in h\_ $\{$2$\}$},}\ }\href@noop {}
  {\bibfield  {journal} {\bibinfo  {journal} {Physical review letters}\
  }\textbf {\bibinfo {volume} {63}},\ \bibinfo {pages} {2040--2043} (\bibinfo
  {year} {1989})}\BibitemShut {NoStop}%
\bibitem [{\citenamefont {Teweldebrhan}\ and\ \citenamefont
  {Balandin}(2009)}]{teweldebrhan2009modification}%
  \BibitemOpen
  \bibfield  {author} {\bibinfo {author} {\bibfnamefont {D.}~\bibnamefont
  {Teweldebrhan}}\ and\ \bibinfo {author} {\bibfnamefont {A.}~\bibnamefont
  {Balandin}},\ }\bibfield  {title} {\enquote {\bibinfo {title} {Modification
  of graphene properties due to electron-beam irradiation},}\ }\href@noop {}
  {\bibfield  {journal} {\bibinfo  {journal} {Applied Physics Letters}\
  }\textbf {\bibinfo {volume} {94}},\ \bibinfo {pages} {013101--013101}
  (\bibinfo {year} {2009})}\BibitemShut {NoStop}%
\bibitem [{\citenamefont {Iqbal}\ \emph {et~al.}(2012)\citenamefont {Iqbal},
  \citenamefont {Kumar~Singh}, \citenamefont {Iqbal}, \citenamefont {Seo},\
  and\ \citenamefont {Eom}}]{iqbal2012effect}%
  \BibitemOpen
  \bibfield  {author} {\bibinfo {author} {\bibfnamefont {M.}~\bibnamefont
  {Iqbal}}, \bibinfo {author} {\bibfnamefont {A.}~\bibnamefont {Kumar~Singh}},
  \bibinfo {author} {\bibfnamefont {M.}~\bibnamefont {Iqbal}}, \bibinfo
  {author} {\bibfnamefont {S.}~\bibnamefont {Seo}}, \ and\ \bibinfo {author}
  {\bibfnamefont {J.}~\bibnamefont {Eom}},\ }\bibfield  {title} {\enquote
  {\bibinfo {title} {Effect of e-beam irradiation on graphene layer grown by
  chemical vapor deposition},}\ }\href@noop {} {\bibfield  {journal} {\bibinfo
  {journal} {Journal of Applied Physics}\ }\textbf {\bibinfo {volume} {111}},\
  \bibinfo {pages} {084307--084307} (\bibinfo {year} {2012})}\BibitemShut
  {NoStop}%
\bibitem [{\citenamefont {Elias}\ \emph {et~al.}(2009)\citenamefont {Elias},
  \citenamefont {Nair}, \citenamefont {Mohiuddin}, \citenamefont {Morozov},
  \citenamefont {Blake}, \citenamefont {Halsall}, \citenamefont {Ferrari},
  \citenamefont {Boukhvalov}, \citenamefont {Katsnelson}, \citenamefont {Geim}
  \emph {et~al.}}]{elias2009control}%
  \BibitemOpen
  \bibfield  {author} {\bibinfo {author} {\bibfnamefont {D.}~\bibnamefont
  {Elias}}, \bibinfo {author} {\bibfnamefont {R.}~\bibnamefont {Nair}},
  \bibinfo {author} {\bibfnamefont {T.}~\bibnamefont {Mohiuddin}}, \bibinfo
  {author} {\bibfnamefont {S.}~\bibnamefont {Morozov}}, \bibinfo {author}
  {\bibfnamefont {P.}~\bibnamefont {Blake}}, \bibinfo {author} {\bibfnamefont
  {M.}~\bibnamefont {Halsall}}, \bibinfo {author} {\bibfnamefont
  {A.}~\bibnamefont {Ferrari}}, \bibinfo {author} {\bibfnamefont
  {D.}~\bibnamefont {Boukhvalov}}, \bibinfo {author} {\bibfnamefont
  {M.}~\bibnamefont {Katsnelson}}, \bibinfo {author} {\bibfnamefont
  {A.}~\bibnamefont {Geim}},  \emph {et~al.},\ }\bibfield  {title} {\enquote
  {\bibinfo {title} {Control of graphene's properties by reversible
  hydrogenation: evidence for graphane},}\ }\href@noop {} {\bibfield  {journal}
  {\bibinfo  {journal} {Science}\ }\textbf {\bibinfo {volume} {323}},\ \bibinfo
  {pages} {610--613} (\bibinfo {year} {2009})}\BibitemShut {NoStop}%
\bibitem [{\citenamefont {Ferrari}\ and\ \citenamefont
  {Robertson}(2000)}]{ferrari2000interpretation}%
  \BibitemOpen
  \bibfield  {author} {\bibinfo {author} {\bibfnamefont {A.}~\bibnamefont
  {Ferrari}}\ and\ \bibinfo {author} {\bibfnamefont {J.}~\bibnamefont
  {Robertson}},\ }\bibfield  {title} {\enquote {\bibinfo {title}
  {Interpretation of raman spectra of disordered and amorphous carbon},}\
  }\href@noop {} {\bibfield  {journal} {\bibinfo  {journal} {Physical Review
  B}\ }\textbf {\bibinfo {volume} {61}},\ \bibinfo {pages} {14095} (\bibinfo
  {year} {2000})}\BibitemShut {NoStop}%
\bibitem [{\citenamefont {Kudin}\ \emph {et~al.}(2008)\citenamefont {Kudin},
  \citenamefont {Ozbas}, \citenamefont {Schniepp}, \citenamefont {Prud'Homme},
  \citenamefont {Aksay},\ and\ \citenamefont {Car}}]{kudin2008raman}%
  \BibitemOpen
  \bibfield  {author} {\bibinfo {author} {\bibfnamefont {K.~N.}\ \bibnamefont
  {Kudin}}, \bibinfo {author} {\bibfnamefont {B.}~\bibnamefont {Ozbas}},
  \bibinfo {author} {\bibfnamefont {H.~C.}\ \bibnamefont {Schniepp}}, \bibinfo
  {author} {\bibfnamefont {R.~K.}\ \bibnamefont {Prud'Homme}}, \bibinfo
  {author} {\bibfnamefont {I.~A.}\ \bibnamefont {Aksay}}, \ and\ \bibinfo
  {author} {\bibfnamefont {R.}~\bibnamefont {Car}},\ }\bibfield  {title}
  {\enquote {\bibinfo {title} {Raman spectra of graphite oxide and
  functionalized graphene sheets},}\ }\href@noop {} {\bibfield  {journal}
  {\bibinfo  {journal} {Nano letters}\ }\textbf {\bibinfo {volume} {8}},\
  \bibinfo {pages} {36--41} (\bibinfo {year} {2008})}\BibitemShut {NoStop}%
\bibitem [{\citenamefont {Scholtz}, \citenamefont {Dijkkamp},\ and\
  \citenamefont {Schmitz}(1996)}]{scholtz1996secondary}%
  \BibitemOpen
  \bibfield  {author} {\bibinfo {author} {\bibfnamefont {J.}~\bibnamefont
  {Scholtz}}, \bibinfo {author} {\bibfnamefont {D.}~\bibnamefont {Dijkkamp}}, \
  and\ \bibinfo {author} {\bibfnamefont {R.}~\bibnamefont {Schmitz}},\
  }\bibfield  {title} {\enquote {\bibinfo {title} {Secondary electron emission
  properties},}\ }\href@noop {} {\bibfield  {journal} {\bibinfo  {journal}
  {Philips journal of research}\ }\textbf {\bibinfo {volume} {50}},\ \bibinfo
  {pages} {375--389} (\bibinfo {year} {1996})}\BibitemShut {NoStop}%
\bibitem [{\citenamefont {Ebinger}, \citenamefont {Yates}\ \emph
  {et~al.}(1998)\citenamefont {Ebinger}, \citenamefont {Yates} \emph
  {et~al.}}]{ebinger1998electron}%
  \BibitemOpen
  \bibfield  {author} {\bibinfo {author} {\bibfnamefont {H.}~\bibnamefont
  {Ebinger}}, \bibinfo {author} {\bibfnamefont {J.}~\bibnamefont {Yates}},
  \emph {et~al.},\ }\bibfield  {title} {\enquote {\bibinfo {title}
  {Electron-impact-induced oxidation of al (111) in water vapor: Relation to
  the cabrera-mott mechanism},}\ }\href@noop {} {\bibfield  {journal} {\bibinfo
   {journal} {Physical Review B}\ }\textbf {\bibinfo {volume} {57}},\ \bibinfo
  {pages} {1976} (\bibinfo {year} {1998})}\BibitemShut {NoStop}%
\bibitem [{\citenamefont {Kaciulis}(2011)}]{kaciulisspectroscopy}%
  \BibitemOpen
  \bibfield  {author} {\bibinfo {author} {\bibfnamefont {S.}~\bibnamefont
  {Kaciulis}},\ }\bibfield  {title} {\enquote {\bibinfo {title} {Spectroscopy
  of carbon: from diamond to nitride films},}\ }\href@noop {} {\bibfield
  {journal} {\bibinfo  {journal} {Surface and Interface Analysis}\ } (\bibinfo
  {year} {2011})}\BibitemShut {NoStop}%
\bibitem [{\citenamefont {Patsalas}\ \emph {et~al.}(2001)\citenamefont
  {Patsalas}, \citenamefont {Handrea}, \citenamefont {Logothetidis},
  \citenamefont {Gioti}, \citenamefont {Kennou},\ and\ \citenamefont
  {Kautek}}]{patsalas2001complementary}%
  \BibitemOpen
  \bibfield  {author} {\bibinfo {author} {\bibfnamefont {P.}~\bibnamefont
  {Patsalas}}, \bibinfo {author} {\bibfnamefont {M.}~\bibnamefont {Handrea}},
  \bibinfo {author} {\bibfnamefont {S.}~\bibnamefont {Logothetidis}}, \bibinfo
  {author} {\bibfnamefont {M.}~\bibnamefont {Gioti}}, \bibinfo {author}
  {\bibfnamefont {S.}~\bibnamefont {Kennou}}, \ and\ \bibinfo {author}
  {\bibfnamefont {W.}~\bibnamefont {Kautek}},\ }\bibfield  {title} {\enquote
  {\bibinfo {title} {A complementary study of bonding and electronic structure
  of amorphous carbon films by electron spectroscopy and optical techniques},}\
  }\href@noop {} {\bibfield  {journal} {\bibinfo  {journal} {Diamond and
  related materials}\ }\textbf {\bibinfo {volume} {10}},\ \bibinfo {pages}
  {960--964} (\bibinfo {year} {2001})}\BibitemShut {NoStop}%
\bibitem [{\citenamefont {Bagri}\ \emph {et~al.}(2010)\citenamefont {Bagri},
  \citenamefont {Mattevi}, \citenamefont {Acik}, \citenamefont {Chabal},
  \citenamefont {Chhowalla},\ and\ \citenamefont
  {Shenoy}}]{bagri2010structural}%
  \BibitemOpen
  \bibfield  {author} {\bibinfo {author} {\bibfnamefont {A.}~\bibnamefont
  {Bagri}}, \bibinfo {author} {\bibfnamefont {C.}~\bibnamefont {Mattevi}},
  \bibinfo {author} {\bibfnamefont {M.}~\bibnamefont {Acik}}, \bibinfo {author}
  {\bibfnamefont {Y.}~\bibnamefont {Chabal}}, \bibinfo {author} {\bibfnamefont
  {M.}~\bibnamefont {Chhowalla}}, \ and\ \bibinfo {author} {\bibfnamefont
  {V.}~\bibnamefont {Shenoy}},\ }\bibfield  {title} {\enquote {\bibinfo {title}
  {Structural evolution during the reduction of chemically derived graphene
  oxide},}\ }\href@noop {} {\bibfield  {journal} {\bibinfo  {journal} {Nature
  chemistry}\ }\textbf {\bibinfo {volume} {2}},\ \bibinfo {pages} {581--587}
  (\bibinfo {year} {2010})}\BibitemShut {NoStop}%
\end{thebibliography}%
\include{euvref.bib}

\end{document}